\documentclass[prd,amsmath,amssymb,reprint,graphicx]{revtex4-1}

\usepackage[pdftex]{graphicx}
\usepackage{dcolumn}
\usepackage{bm}

\draft 

\begin{document}


\title{Energy deposition on nuclear emulsion by slow recoil ions for directional dark matter searches} 



\author{Akira Hitachi}
 \email[]{a.hitachi@kurenai.waseda.jp}
 \affiliation{Research Institute for Science and Engineering, Waseda University, Shinjuku, Tokyo 169-8555, Japan}
\author{A. Mozumder}
 \affiliation{Radiation Laboratory, University of Notre dame, Notre Dame, IN 46556-5674, USA}
\author{Kiseki D. Nakamura}
 \email[]{kiseki@harbor.kobe-u.ac.jp}
 \affiliation{Particle Physics Laboratory, Kobe University, Kobe, Hyogo 657-8501, Japan}

\date{\today}

\begin{abstract}
	The electronic energy deposited on nuclear emulsions due to C ions of 5 -- 200~keV and Kr ions of 5 -- 600~keV are evaluated and compared with those due to fast ions for design and construction of fine grain nuclear emulsion for directional dark matter searches.
	Nuclear quenching factors and the electronic LET (linear energy transfer), the specific electronic energy deposited along the ion track, are evaluated.
	The so-called core and penumbra of heavy-ion track structure is modified for understanding the track due to recoil ions produced by dark matter candidate, WIMPs, striking nucleus in the AgBr crystal of nuclear emulsion.
	The very heavy recoil ions, 100 -- 180~keV Pb ions, produced in $\alpha$-decay are also studied.
	In addition, the track structures due to proton ions of 25 -- 80~keV are evaluated to consider the influence of background neutrons in underground laboratories. 
\end{abstract}


\maketitle 

\section{Introduceion}
Identification of dark matter is one of compelling challenges in cosmology, astrophysics and particle physics.
The scientific evidence, such as the rotational velocity of galaxies in the cluster \cite{Zwickey1933}, the rotational velocity of stars and gases in the galaxy \cite{rotation_curve,rotation_curve2} and the gravitational lensing \cite{weak_lensing,weak_lensing2,weak_lensing3} confirm the existence of nonbaryonic dark matter.
The unseen dark matter accounts for a quarter of the universe energy.
The Galaxy is surrounded by dark matter.
The solar system is traveling around the galactic center at 230~km/s towards Cygnus.
The detection of WIMPs, Weakly Interacting Massive Particles, the leading candidates for galactic dark matter, usually observe the ionization, excitation and chemical reactions caused by recoil ions of a few to a few tens of keV energy, produced by elastic scattering with WIMPs \cite{DM_search,DM_search2,DM_search3,DM_search4,DM_search5}.
Various types of detectors are in operation observing two or more kinds of signals exploiting deference in the response for slow recoil ions and background $\gamma$-rays \cite{XENON, PICO, CDMS}. 
The resulting kinetic spectrum for recoil ions will not be monochromatic but may be described as similar to exponential.
The directional detection of WIMPs will give excellent capability of discriminating nuclear recoil signals from background $\gamma$-rays and neutrons, etc., by exploiting daily modulation of WIMP wind \cite{DM_search5,DM_search6}. 
\par
Nuclear emulsions have been used for detecting various particles with wide ranges of energy.
A careful analysis of the details of the track gives much information about the type of particle and its energy \cite{emulsion_JRE}.
Nuclear emulsions have exceptional capabilities for adjusting sensitivity for particle and its energy, according to the experimental purpose, by changing the grain size and the developing procedure, etc.
Most of $\gamma$-rays, which is the major source of the backgrounds, can be rejected by adjusting the sensitivity of emulsion.
The fine grain nuclear emulsions have been proposed and being developed for the directional detection of WIMPs \cite{emulsion_NATSUME,emulsion_NAKA,emulsion_NDA}. 
The basic information for the interaction of slow ions ($v < v_0$, where $v$ is the velocity of the ion and $v_0 \approx c/137 = 2.2\times10^8~\rm{cm/sec}$ is the Bohr velocity, and $c$ is the velocity of the light) with detector media such as stopping power, energy sharing, and quenching are crucial for design and construction of dark matter detectors.
The electronic energy deposited on the nuclear emulsions is discussed for understanding the track images.
We propose a simple model to predict the footprints of WIMPs in nuclear emulsions.
The electronic LET (linear energy transfer) plays an important role in slow ion collisions \cite{DM_search4,Hitachi2005,Hitachi2008}.
\par
The nuclear emulsion consists of AgBr crystal (grain) sustained in gelatin \cite{emulsion_NATSUME,emulsion_NAKA,emulsion_NDA}.
The conduction electrons created by charged particles may become trapped, combined with mobile silver ions and form aggregates of silver atoms.
The latent image specks are formed on each crystal by following reactions:
\begin{eqnarray}
&&{\rm e}^- + {\rm Ag}^+ \rightarrow {\rm Ag} \\
&&{\rm e}^- + {\rm Ag}^+ + {\rm Ag} \rightarrow {\rm Ag}_2 \cdots {\rm Ag}_n.
\end{eqnarray}
Development makes Ag filament structure.
The density of AgBr crystal is 6.473~g/cm$^3$ and the number density of AgBr is $2\times 10^{22}~\rm cm^{-3}$.
The atomic distance $a$ is 2.88~\AA.
The direct band gap energy $E_{\rm g}$ is 4.292~eV \cite{Carrera1971,Testa1988} and the average energy $W$ required for an ionization is 5.8~eV \cite{Yamakawa1951}.
The crystal size of the grain for standard emulsion is 200~nm and that for fine grain emulsion is 18 -- 40~nm.
The density of fine-grain emulsion is 3.2~g/cm$^3$ and the mass ratio of the atoms can be approximated as Ag:Br:C(N,O) = 9:7:2.
The number density of atoms is $8\times 10^{22}~{\rm cm}^{-3}$.
The sensitivity of the emulsion depends very much on the grain size and also on the developer, etc.

\section{Heavy ion track}
It may be better to take a simple model to obtain some idea of response of emulsion to recoil ions because composition and structure and chemical reactions in nuclear emulsion are quite complicated.
In addition, accurate values of some physical and chemical quantities are hard to obtain.
The $\delta$-ray theory of track structure has been proposed for the response of nuclear emulsions \cite{Kats1975,MPR1986}, since the “visual” radial image of the track is largely determined by the penetration of energetic secondary electrons ($\delta$-rays).
However, the $\delta$-ray theory is not suitable for slow recoil ions since the energy and the range of $\delta$-rays are too small.
We propose a model close to the so-called core and penumbra of heavy-ion track structure.
The track for fast ions such as $\alpha$-particles and ions of several MeV/n to a few GeV/n can be regarded as cylindrical geometry, consists of the high-density core and surrounding less dense penumbra \cite{Mozumder1999,Chatterjee1976,Magee1980}.
The core is due to glancing (distant) collisions and the penumbra is formed by $\delta$-rays produced by knock-on (close) collisions.
Glancing collisions transfer small amount of energy frequently within the core, a region of finite size.
The radius of the core $r_{\rm c}$ is given by Bohr-criterion, 
\begin{equation}
r_{\rm B}=\hbar v/2E_1,
\end{equation}
where $\hbar$ is Plank's constant divided by $2\pi$, $v$ is the velocity of incident ion and $E_1$ is the energy of lowest electronic excited state of the medium.
The stopping power theory supports the equipartition of the total energy loss between the glancing and close collisions.
The initial radial distribution of track core given by glancing collisions may be approximated as Gaussian with a size parameter $a_0$ as the same as the core radius \cite{Hitachi1992}
\begin{equation}\label{eq_Dg}
D_{\rm g}=\frac{LET/2}{\pi a_0^2}\exp{(-r^2/a_0^2)},
\end{equation}
where $r$ is the radius and $LET$ is the linear energy transfer.
\par
\begin{figure}[t]
	\includegraphics[width=0.9\linewidth]{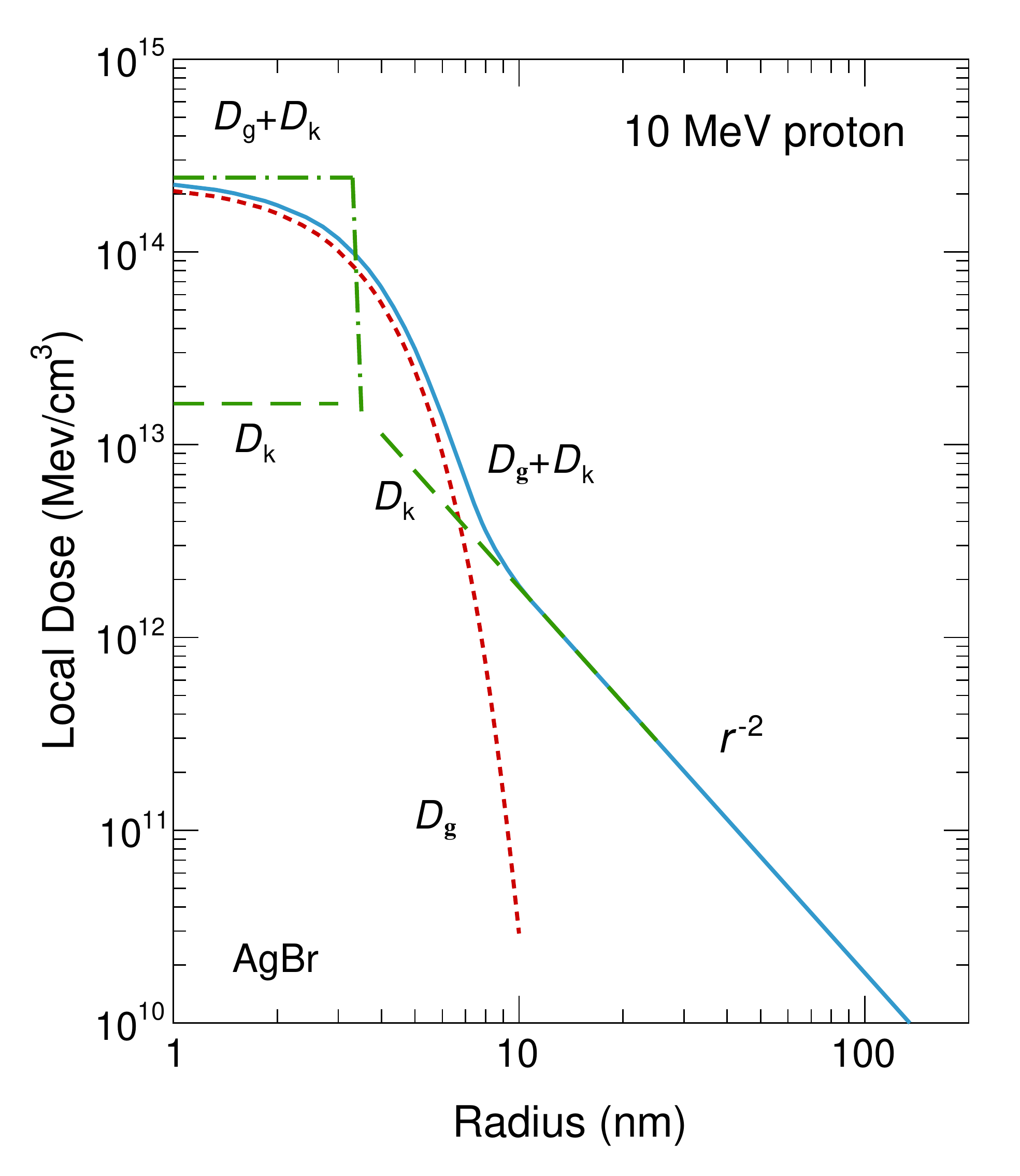}
	\caption{
		The so-called core and penumbra of heavy-ion track structure shown for 10~MeV protons in AgBr crystal.
		Solid and dotted curves show the present model.
		Dashed and dot-dashed lines show the sample model commonly used \cite{Chatterjee1976,Magee1980}.
		Only a part of penumbra is shown.
		Suffix g and $\mathfrak{g}$ are for glancing collisions, and k is for knock-on collisions.
	}
	\label{fig_1}
\end{figure}

Knock-on collisions transfer large amount of energy less frequently producing $\delta$-rays.
The $\delta$-rays deposit a part of its energy in the core and the rest in surrounding penumbra.
With a simple model, the $\delta$-ray contributions for the core and the penumbra are written as \cite{Chatterjee1976,Magee1980},
\begin{eqnarray}
\label{eq_Dk1} &&D_{\rm k}=\frac{LET/2}{2\pi r_{\rm c}^2\ln{(\sqrt{\rm e}r_{\rm p}/r_{\rm c})}},\qquad r\le r_{\rm c} \\
\label{eq_Dk2} &&D_{\rm k}=\frac{LET/2}{2\pi r^2\ln{(\sqrt{\rm e}r_{\rm p}/r_{\rm c})}},\qquad r_{\rm c}<r\le r_{\rm p},
\end{eqnarray}
where $r_{\rm p}$ is the radius of the penumbra and is given by the range of $\delta$-rays of the maximum energy.
Normal ejection of $\delta$-rays with constant energy loss is assumed.
The kinematically limited maximum $\delta$-ray energy is given by $4Em_{\rm e}/M$ where $m_{\rm e}$ is the electron mass and $M$ is the mass of incident particles, or $2m_{\rm e}c^2\beta^2/(1-\beta^2)$ where $\beta=v/c$ when the incident ion is relativistic. A typical initial radial distribution of dose in AgBr crystal due to 10~MeV proton is shown in Fig.~\ref{fig_1}.
The direct energy gap $E_{\rm g}$ was used for $E_1$.

\begin{table*}[tb]
	\caption{\label{tab:table1}
		Data for various particle tracks in AgBr crystal.
		The ranges are projected range in emulsion \cite{emulsion_NATSUME,SRIM2010}.
		Values for $r_{\rm c}$ and $r_{\rm p}$ are at the initial energy $E$.
		$r_{\rm c}$ values for low energy p, C, Kr and Pb ions show expanded core radii, see the text. 
	}
	\begin{ruledtabular}
		\begin{tabular}{cccccccc}
			particles &energy  &range   &$q_{\rm nc}$  &$\langle LET_{\rm el}\rangle$ &$E_{\delta}^{\rm max}$ &$r_{\rm c}$ &$r_{\rm p}$ \\
			          &keV     &$\mu$m  &$E_{\eta}/E$  &keV/$\mu$m                    &keV                    &nm          &nm          \\
			\hline
			protons   &$\ \,$25          &0.29$\ \,$ &1.0   &$\ \,$87 &0.05  &0.48\footnotemark[1] &-\\
			protons   &$\ \,$80          &0.74$\ \,$ &1.0   &109      &0.18  &0.46\footnotemark[1] &1.5\\
			protons   &10$\times 10^3$   &630        &1.0   &$\ \,$16 &22    &3.3                  &2,070\\
			alphas    &5.3$\times 10^3$  &25         &1.0   &212      &2.9   &1.2                  &63\\
			C         &3.48$\times 10^6$ &-          &1.0   &$\ \,$50 &735   &11                   &5.9$\times 10^5$\\
			C         &$\ \,$30          &0.093      &0.58  &190      &-     &0.7\footnotemark[1]  &-\\
			C         &100               &0.30$\ \,$ &0.77  &260      &-     &0.8\footnotemark[1]  &-\\
			Kr        &$\ \,$30          &0.024      &0.27  &340      &-     &0.9\footnotemark[1]  &-\\
			Kr        &100               &0.060      &0.33  &550      &-     &1.2\footnotemark[1]  &-\\
			Kr        &200               &0.11$\ \,$ &0.37  &680      &-     &1.3\footnotemark[1]  &-\\
			Kr        &600               &0.35$\ \,$ &0.48  &830      &-     &1.4\footnotemark[1]  &-\\
			Pb        &100               &0.040      &0.14  &360      &-     &1.0\footnotemark[1]  &-\\
			Pb        &170               &0.058      &0.19  &540      &-     &1.2\footnotemark[1]  &-\\
		\end{tabular}
	\end{ruledtabular}
	\footnotetext[1]{Expanded core radius}
\end{table*}

\par
The simple model commonly used \cite{Chatterjee1976} assumes also constant dose due to glancing collisions, instead of Eq.~(\ref{eq_Dg})
\begin{equation}\label{eq_Dg2}
D_{\mathfrak{g}}=\frac{LET/2}{2\pi r_{\rm c}^2},\qquad r\le r_{\rm c}.
\end{equation}
Then, the dose within the core is the sum of Eqs.~(\ref{eq_Dk1}) and (\ref{eq_Dg2}) as shown in Fig.~\ref{fig_1}.
The sharp distinction made in the model between core and penumbra is an artificial concept introduced mainly for conducting the analytical process \cite{Chatterjee1976}.
Introducing a Gaussian distribution for the glancing collisions, Eq.~(\ref{eq_Dg}), the present model makes the distinction moderate.
The Gaussian form can be used for treating redistribution of energy and chemical reactions in track core, e.g., diffusion reactions in scintillation quenching in liquid Ar \cite{Hitachi1992}.
The fraction of energy deposited with in a cylinder of radius $r$ is given by, 
\begin{equation}\label{eq_F}
\begin{split}
F=\frac{1}{2}+\frac{1}{4\ln{(\sqrt{\rm e}r_{\rm p}/r_{\rm c})}} +\frac{\ln{(r/r_{\rm c})}}{2\ln{(\sqrt{\rm e}r_{\rm p}/r_{\rm c})}}, \\ \qquad r_{\rm c}<r\le r_{\rm p}.
\end{split}
\end{equation}
The first term represents a contribution of glancing collisions and the second term is due to $\delta$-rays to the core.
The third term is due to $\delta$-rays deposited in the penumbra with in radius $r$.
Eq.~(\ref{eq_F}) can be also applied for the present model when $r$ is large enough than $r_{\rm c}$.
Values for track parameters of various particles are listed in Table~\ref{tab:table1}.

\par
For relativistic particles, Fermi's criterion
\begin{equation}\label{eq_Fermi}
r_{\rm F}=\lambda\beta,
\end{equation}
is used for $r_{\rm c}$, where $\lambda$ is the maximum core size and is given by \cite{Mozumder1974}, 
\begin{equation}\label{eq_lambda}
\lambda=\chi_{\rm max}c/n\omega_0,
\end{equation}
where $\chi_{\rm max}=1.074$ and $n = \epsilon^{1/2}$ is the refractive index, taken to be 2.253 for AgBr crystal.
The angular frequency $\omega_0$ in the compound consist of light elements, such as water, the geometrical mean ionization potential of the electrons (excluding the K electrons) is used. 
\par
Often, the following form \cite{Magee1980}
\begin{equation}\label{eq_lambda2}
\lambda=c/\omega_{\rm p}
\end{equation}
is used with the plasma frequency,
\begin{equation}\label{eq_omega}
\omega_{\rm p}=\sqrt{\frac{n_{\rm e}e^2}{m^{*}\epsilon_0}},
\end{equation}
where $n_{\rm e}$ is the number density of electrons, $e$ is the charge of electron, $m^{*}$ is the effective mass of the electron, and $\epsilon_0$ is the permittivity of free space.
The values for $\lambda$ in water are 93~\AA \cite{Mozumder1974} and 103~\AA \cite{Chatterjee1976}, respectively, for Eqs.~(\ref{eq_lambda}) and (\ref{eq_omega}), and are close to each other.
However, the use of the plasma frequency for heavy elements such as AgBr gives very large $\hbar\omega_0$, and consequently, unreasonably small $\lambda$ value.
Because, AgBr crystal has the band structure, we take the average energy $W$ required for produce a hole-electron pair which gives $\hbar\omega_0$ = 5.8~eV or that $\omega_0 = 8.81\times 10^{15}\,{\rm sec^{-1}}$.
Then we obtain $\lambda$ = 170~\AA with Eq.~(\ref{eq_lambda}) in AgBr crystal.
\par
The track dimensions $r_{\rm c}$ and $r_{\rm p}$ depend only on the particle velocity $\beta$.
$D_{\rm k}$ decrease as $r^2$ at a large $r$.
The same shape can be applied for various ions at the same energy per nucleon, MeV/n.
However, dose (energy density) depends on the LET and thus on the particle charge.
LET scales as square of the effective charge, $Z_{\rm eff}^2$, which is a function of the velocity \cite{Magee1980}.

\section{Slow recoil ions}
\subsection{Stopping powers}
For the interaction of slow ions with matter, the nuclear stopping power $S_{\rm n}$ is of the same order of magnitude as the electronic stopping power $S_{\rm e}$ \cite{Lindhard1963}.
The total stopping power $S_{\rm T}$ is the sum of the two; $S_{\rm T} = S_{\rm n} + S_{\rm e}$.
The value for $r_{\rm c}$ given by the Bohr criterion becomes unreasonably small to make excitation higher than $E_1$ by slow ions.
The projectile cannot come close enough to the target atom due to repulsive potential in ordinary manner.
In addition, even the kinematically limited maximum energy for secondary electrons may not exceed $E_1$ for some cases.
These means that usual theories for $S_{\rm e}$ based on ion-atom collisions such as Bohr's classical theory and Bethe's quantum mechanical theory are inaplicable to those slow collisions.
Lindhard et al. considered dielectric response \cite{Lindhard1964}.
A charged particle incident on the electron gas causes polarization and changes the dielectric constant.
Consequently, the incident particle receives the electric force opposite direction that causes $S_{\rm e}$.
For slow ions, $S_{\rm e}$ is expressed as $({\rm d}\varepsilon/{\rm d}\rho)_{\rm e} \approx k\varepsilon^{1/2}$ where $\varepsilon$ is the dimensionless energy and $\rho$ is the dimensionless range.
Based on the Thomas-Fermi treatment, $S_{\rm e}$ is given to a first approximation by \cite{Lindhard1961},
\begin{equation}\label{eq_Se}
\begin{split}
S_{\rm e}=\xi_{\rm e}\times 8\pi e^2a_{\rm B} \frac{Z_1Z_2}{(Z_1^{2/3}+Z_2^{2/3})^{3/2}} \frac{v}{v_0}, \\ {\rm with\,} \xi_{\rm e} \approx Z_1^{1/6},
\end{split}
\end{equation}
where $Z$ and $A$ the atomic number and the atomic mass and suffix 1 and 2 are for projectile and the target, respectively, and $a_{\rm B} = \hbar^2/m_{\rm e}e^2$ = 0.529 \AA $\,$ is the Bohr radius.
The parameter $k$ is given by Eq.~(\ref{eq_Se}) and is expressed as $k = 0.133Z_2^{2/3}A_2^{-1/2}$ for $Z_1=Z_2$.
For most cases, $k = 0.1 \sim 0.2$.
\par
The nuclear process follows the usual procedure of a screened Rutherford scattering.
The nuclear stopping power can be expressed by the analytical expression using the Firsov potential \cite{Biersack1968},
\begin{equation}\label{eq_Sn}
S_{\rm n}=\frac{4\pi a_{\rm TFF}A_1Z_1Z_2e^2}{A_1+A_2} \frac{\ln{\varepsilon}}{2\varepsilon(1-\varepsilon^C)},
\end{equation}
where $C = -1.49$ and $a_{\rm TFF}$ is the Thomas-Fermi-Firsov screening radius,
\begin{equation}\label{eq_aTFF}
a_{\rm TFF}=0.8853a_{\rm B}/(Z_1^{1/2}+Z_2^{1/2})^{2/3}.
\end{equation}
The stopping powers discussed above give the same values as those in the HMI tables \cite{Biersack1975} at a low $E$.
The energy is converted to $\varepsilon$ by
\begin{equation}\label{eq_epsilon}
\varepsilon=C_{\varepsilon}E=\frac{a_{\rm TFF}A_2}{Z_1Z_2e^2(A_1+A_2)}E.
\end{equation}
Eq.~(\ref{eq_epsilon}) becomes $\varepsilon = 11.5Z_2^{-7/3}E$ for $Z_1=Z_2$.
The values of $C_{\varepsilon}$ are with $E$ in keV, 0.1759, 0.0350 and 0.00268, 0.00500 for C ions in C, C in Kr, Kr ions in C, and Kr ions in Kr, respectively. 
\par
The effect of the charge state $Q$ of the projectile on the stopping power is determined by the screening radius ($a_{\rm TFF}$ in Eq.~(\ref{eq_aTFF}) or the corresponding part in Eq.~(\ref{eq_Se})) \cite{Tilinin1995}.
The dependence on $Q$ is moderate as it can be seen by replacing $Z_1$ by $\xi_1 = Z_1-Q$.
The projectiles of various $Q$ exchange electrons with target atoms and soon become the charge equilibrium according to the velocity.
Values of $Q$ determined by the charge equilibrium for slow ions are small, therefore $Q$ affects quite weakly the stopping power.
If the different charge states give different results, it is likely through the surface effects.

\begin{figure}[t]
	\includegraphics[width=0.9\linewidth,bb=50 510 400 800]{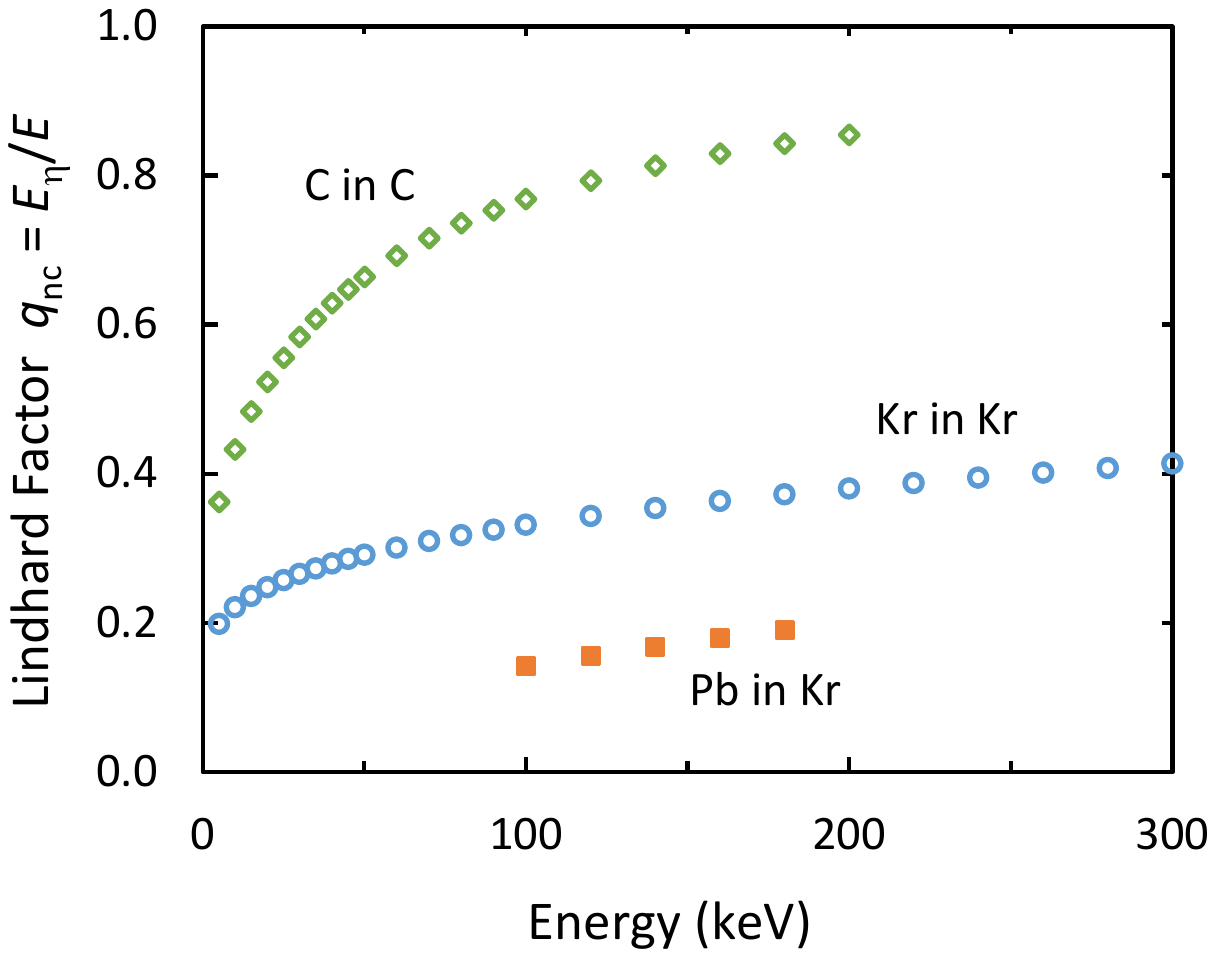}
	\caption{
		The Lindhard factor ($q_{\rm nc} = \eta/\varepsilon$) for C ions in C and Kr ions in Kr as functions of energy.
		The $q_{\rm nc}$ values for Pb ions in Kr are also shown.
	}
	\label{fig_2}
\end{figure}

\subsection{Electronic LET}
The electronic stopping power $S_{\rm e}$ do not directly give the electronic energy deposited to the target matter.
The secondary ions will go into the collisional processes again, and so on.
After cascade processes of stopping collisions, considerable amounts of energy $\nu$ go into atomic motion which is wasted as heat in ordinary detectors.
Only a part of the energy $\eta$ goes to the electronic excitation which can causes ionization, excitation, and chemical reactions.
It is necessary to obtain of the ratio, $q_{\rm nc} = \eta/\varepsilon$ (the nuclear quenching factor or the Lindhard factor) to evaluate the detection efficiency, etc.
Lindard et al. \cite{Lindhard1963} solved the homogeneous integral equation for $\nu\, (=\varepsilon-\eta)$ and gave numerical results for $Z_1=Z_2$ for $k$ = 0.1, 0.15 and 0.2.
The value of $k$ for Kr ions in Kr is 0.158, therefore it can be approximated as $k$ = 0.15.
Following expressions was taken from Fig.~3 in Ref.~\cite{Lindhard1963}:
\begin{equation}\label{eq_eta}
\eta=0.187\varepsilon^{1.362}+0.246\varepsilon^{1.101},\qquad \varepsilon\le 2.
\end{equation}
They also gave a comprehensive formula for $\nu$.
$\eta=\varepsilon-\nu$ is expressed as,
\begin{equation}\label{eq_eta2}
\eta=\frac{k\varepsilon{\textsl g}(\varepsilon)}{1+k\cdot {\textsl g}(\varepsilon)},
\end{equation}
for the $k$ values of $0.1 \sim 0.2$.
The comprehensive formula reproduces the numerical $\nu$ within an accuracy of several \%.
A function ${\textsl g}(\varepsilon)$ is later fitted by Lewin and Smith \cite{DM_search3} as
\begin{equation}\label{eq_ge}
{\textsl g}(\varepsilon)=3\varepsilon^{0.15}+0.7\varepsilon^{0.6}+\varepsilon.	
\end{equation}
Then, the nuclear quenching factor for recoil ions in a single element material, $Z_1=Z_2$, is obtained by interpolation of the numerical results (Figs.~3 and 4 in Ref.~\cite{Lindhard1963}) or the asymptotic form using Eqs.~(\ref{eq_eta2}) and (\ref{eq_ge}) as shown in Fig.~\ref{fig_2}.
\par
\begin{figure}[t]
	\includegraphics[width=0.9\linewidth,bb=50 510 420 800]{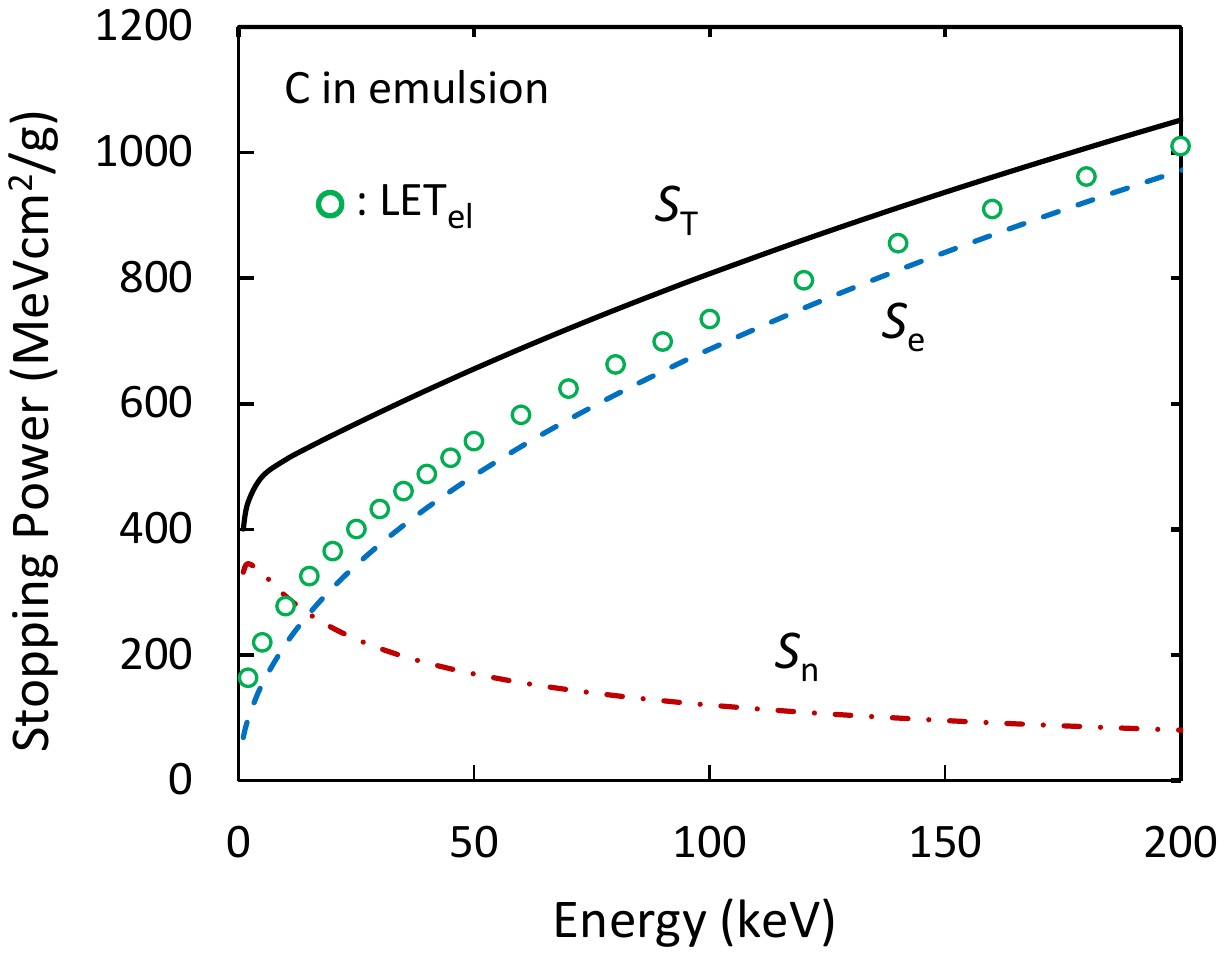}
	\caption{
		The stopping powers, $S_{\rm T}$, $S_{\rm n}$, $S_{\rm e}$, and the electronic LET, $LET_{\rm el}$ ($=-{\rm d}E_{\eta}/{\rm d}x$), for C ions in emulsion as functions of energy.
		$LET_{\rm el}$ is obtained using $q_{\rm nc}$ values for C ions in C, see the text.
	}
	\label{fig_3}
\end{figure}
\begin{figure}[t]
	\includegraphics[width=0.9\linewidth,bb=50 430 420 800]{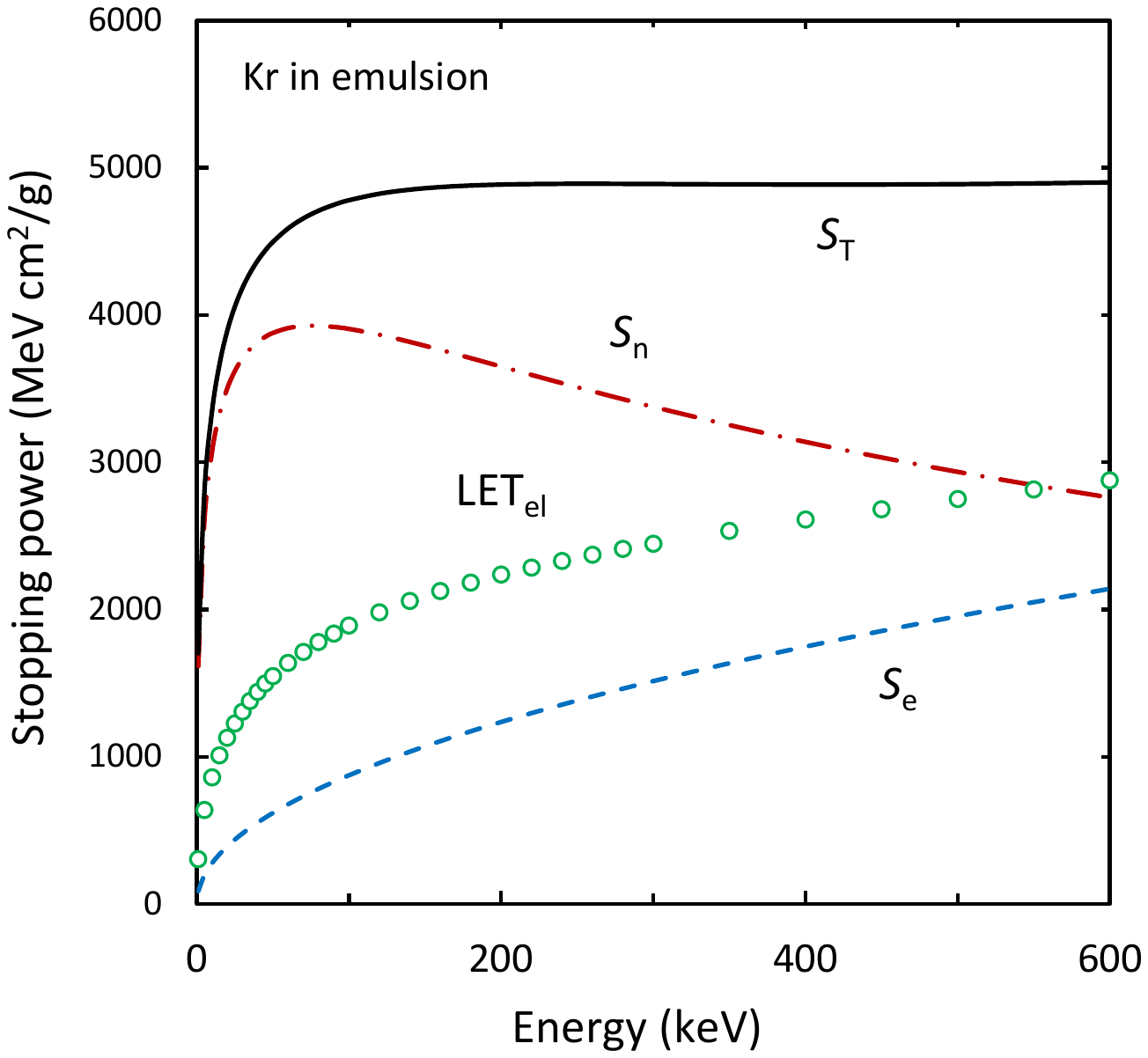}
	\caption{
		The stopping powers, $S_{\rm T}$, $S_{\rm n}$, $S_{\rm e}$, and the electronic LET, $LET_{\rm el}$ ($=-{\rm d}E_{\eta}/{\rm d}x$), for Kr ions in emulsion as functions of energy.
	}
	\label{fig_4}
\end{figure}

The information of microscopic electronic energy deposition is required for evaluate the latent images produced in the nuclear emulsion.
The electronic LET ($LET_{\rm el}$) becomes an important concept in slow ion collisions \cite{Hitachi2005,Hitachi2008}.
We have simply $LET_{\rm el} = - {\rm d}(q_{\rm nc}E)/{\rm d}x$.
However, a little complication comes since $q_{\rm nc}$ (or $\eta$) is an integrated quantity.
Then, we have
\begin{equation}\label{eq_LETel}
\begin{split}
LET_{\rm el}&=-\frac{{\rm d}E_\eta}{{\rm d}x}=-\frac{{\rm d}E_\eta}{{\rm d}E} \frac{{\rm d}E}{{\rm d}x}=\frac{{\rm d}E_\eta}{{\rm d}E} S_{\rm T} \\ &\approx\frac{\Delta E_\eta}{\Delta E} S_{\rm T},
\end{split}
\end{equation}
where $E_\eta = \eta/C_{\varepsilon}$.
An averaged form would make it clear, 
\begin{equation}\label{eq_LETavg}
\langle LET_{\rm el}\rangle=-E_{\eta}/R=-q_{\rm nc}E/R,
\end{equation}
where $R$ is the range.
The electronic LET represents the specific electronic energy deposited along the ion track and is not the same as the electronic stopping power $S_{\rm e}$ as shown in Figs.~\ref{fig_3} and \ref{fig_4}.
$LET_{\rm el}$ is larger than $S_{\rm e}$ for slow ions because secondary ions can give its energy to the electronic excitation when their energy is large enough.
For fast particles, the contributions of nuclear scattering are negligible, therefore, $LET$ and $LET_{\rm el}$ are the same.

\subsection{Slow recoil ion track}
The track structure for slow recoil ions is different from the so-called core and penumbra of heavy-ion track structure discussed above.
We assume most $\delta$-rays produced by recoil ions do not have sufficient energy to effectively escape the core and form an undifferentiated core.
Then, the radial distribution may be approximated as a single Gaussian and $LET/2$ in Eq.~(\ref{eq_Dg}) is replaced by $LET_{\rm el}$ for recoil ions.
For recoil ions, $r_{\rm B}$ becomes less than the interatomic distance $a$, in which case $a$ is taken for $r_{\rm c}$.
The excitation density can be so high that the number density of ionization $n_{\rm i}$ estimated can exceed the number density $n_0$ of AgBr.
When this should occur, redistribution of energy and core expansion may take place, $a_0$ is determined so that $n_{\rm i}$ does not exceed $n_0$.
The maximum local dose $D_{\rm max}$ was set to be $n_{\rm i}\cdot W = 2.08\times 10^{22}\,{\rm cm}^{-3}\times 5.8\,{\rm eV} = 1.20\times 10^{17}\,{\rm MeV/cm^3}$.
\par
The range of electron below 10~keV energy is not reliable because of experimental difficulties.
The Bethe theory becomes invalid in this energy region.
However, in connection with the study of track structure it is of some importance to describe the behavior of $\delta$-rays of a few keV.
Iskef et al.~\cite{Iskef1983} have studied and compared published experimental information on penetration depths of electrons.
They gave following 'best fit' expression applicable to all media between 20~eV to 10~keV energy with a simple scaling factor $Z/A$.
Extrapolated ranges, $R_{\rm ex}$ (in $\mu \rm g/cm^2$), are given by,
\begin{equation}\label{eq_Rex}
\begin{split}
\ln{[(Z/A)R_{\rm ex}]}=&-4.5467+0.31104\ln{E} \\ &+0.7773(\ln{E})^2,
\end{split}
\end{equation}
where the energy $E$ is in eV.
Since, $Z/A$ values for Ag and Br are practically the same, $R_{\rm ex}$ values for Ag was calculated and the range in AgBr crystal was obtained by using the density for AgBr crystal.

\subsection{Very heavy recoil ions in $\alpha$-decay}\label{sec_very_heavy_recoil_ions_in_alpha_decay}
In $\alpha$-decay, the daughter nucleus, such as Pb and Tl, are recoiled with typically 100 -- 170~keV energy.
The very heavy recoil ions in $\alpha$-decay produce WIMP-like signals in detector media, and their contribution to the background signal can be very serious.
It is important to know what signal will be produced.
Lindhard et al. \cite{Lindhard1963} gave a power law approximation for $q_{\rm nc}$ for $Z_1\neq Z_2$ at very low energy.
The model has been applied for binary gases and gave satisfactory results except in hydrocarbons \cite{Hitachi2008}.
We have 
\begin{equation}\label{eq_Eeta}
E_{\eta}=0.0142E^{3/2},
\end{equation}
for Pb ions in Kr (AgBr).

\subsection{Compounds}
The chemical composition of nuclear emulsion is quite complicated and the structure is also not homogeneous.
We assume that only the energy deposited in AgBr crystal is used for the image production and no transfer of energy from gelatin to AgBr crystal.
Composition (the ratio of number densities) of nuclear emulsion are assumed to be Ag:Br:C(N,O) = 0.4:0.4:2.
Light elements such as C, N and O are regarded as C.
H is ignored in stopping calculation except for in the density.
The densities are taken as 6.473 g/cm$^3$ and 3.2 g/cm$^3$, for AgBr crystal and fine-grain nuclear emulsion, respectively.
The stopping powers for slow ions are obtained using Eq. (\ref{eq_Se}) and Eq. (\ref{eq_Sn}) for $S_{\rm e}$ and $S_{\rm n}$, respectively, unless otherwise mentioned.
The stopping powers $S_{\rm mix}$ in compounds are obtained using the Bragg rule, 
\begin{equation}\label{eq_Smix}
S_{\rm mix}=\Sigma N_{\rm i}S_{\rm i}/\Sigma N_{\rm i},
\end{equation}
where $S_{\rm i}$ and $N_{\rm i}$ are the stopping power (in $\rm eV\cdot $\AA$\rm ^2/Atom$) and the number density, respectively, for $i$-th element in target medium. 
\par
The evaluation of $q_{\rm nc}$ for $Z_1\neq Z_2$ is quite hard.
Ag and Br recoil ions are produced in AgBr crystal.
For further simplicity, heavy elements, Ag and Br are regarded as Kr to obtain values of $q_{\rm nc}$ in AgBr crystal, because $Z$ and $A$ for those elements are close to those for Kr.
Light ions such as C, N and O ions, on the other hand, produced in gelatin and may get into AgBr crystal.
$Z$ and $A$ of those light ions are much smaller than those for Ag and Br atoms, therefore it cannot be regarded that the projectile and the target are the same.
A different approach is necessary.
The electronic-to-total stopping power ratio $S_{\rm e}/S_{\rm T}$ for C ions in C differs less than 5\% from that for C ions in Br for $E\geq 20$~keV.
Therefore, it may be safe to take $q_{\rm nc}$ values for C ions in C instead of those for C ions in emulsion except for extremely low energy.
The value of $\eta$ for C ions in C was obtained by Eq. (\ref{eq_eta2}) with $k=0.127$.
It should be regarded as upper limits since it over estimates contributions of the secondary ions to $q_{\rm nc}$.
For H ions, it may be safe to take $q_{\rm nc}$ = 1 in a first approximation.

\section{Results and Discussion}
\subsection{Stopping powers and electronic LET}
The Lindhard factors, $q_{\rm nc}$, for 5 -- 200~keV C ions in C and 5 -- 600~keV Kr ions in Kr are shown in Fig.~\ref{fig_2}.
Values for C ions in C and Kr ions in Kr increases rapidly at a low energy and tend to saturate as energy increases.
The stopping powers and $LET_{\rm el}$ for C and Kr ions in emulsion as a function of energy are shown in Figs.~\ref{fig_3} and \ref{fig_4}.
$S_{\rm e}$ and $LET_{\rm el}$ differ very much in low energy region, where $S_{\rm n}$ is larger than $S_{\rm e}$.
$LET_{\rm el}$ comes close to $S_{\rm e}$ for C ions above $\sim$50~keV.
Since $S_{\rm e} < LET_{\rm el} < S_{\rm T}$, and the most contribution to $S_{\rm T}$ comes from $S_{\rm e}$, it can be seen that the approximation of regarding C ions in emulsion as C ions in C for estimating $q_{\rm nc}$ is verified.
$S_{\rm e}$ and $LET_{\rm el}$ for Kr ions in emulsion differ very much in low energy region and they are still considerably different each other even at 600~keV.

\begin{figure}[t]
	\includegraphics[width=0.9\linewidth,bb=50 480 430 770]{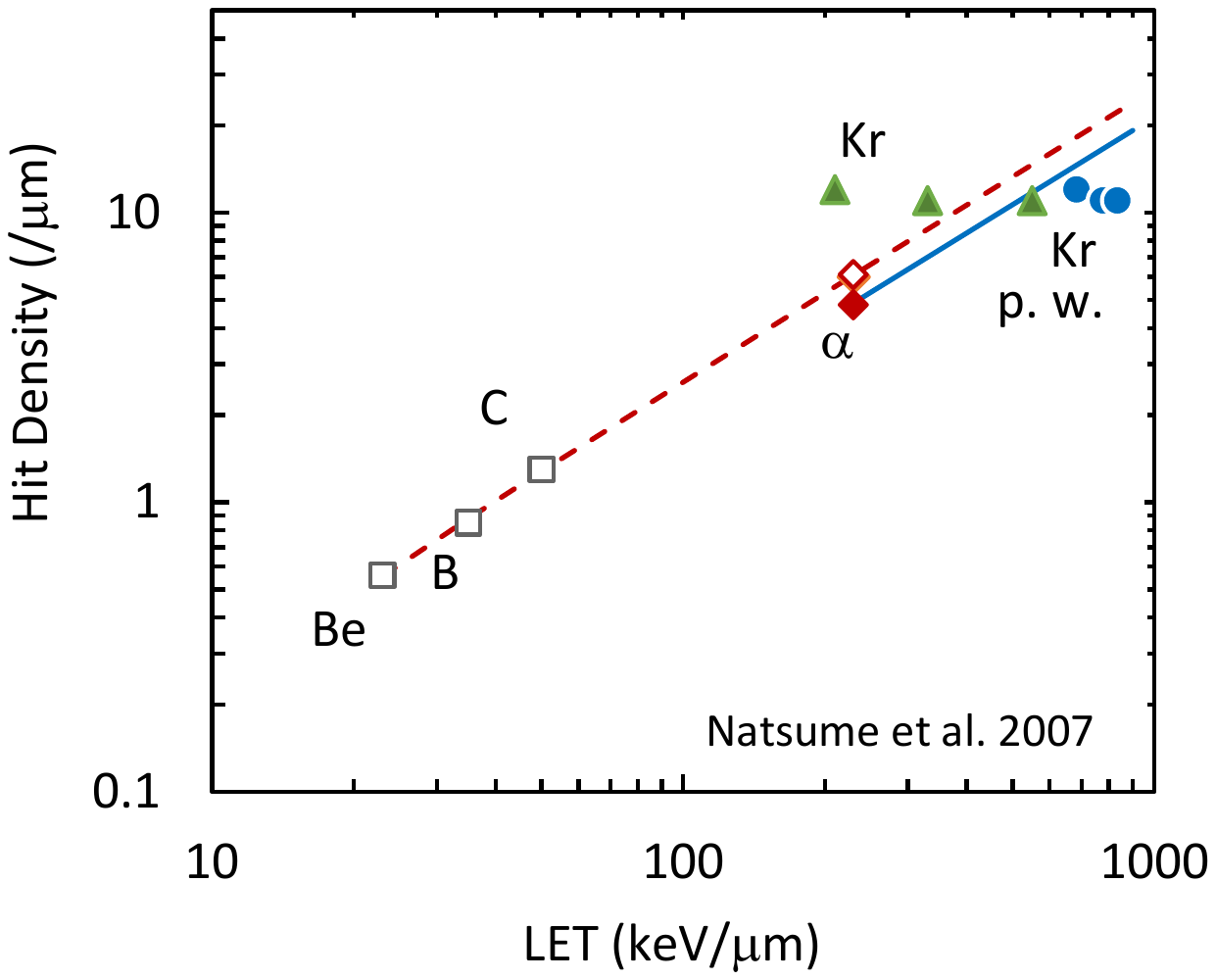}
	\caption{
		The mean hit density for 5.3~MeV $\alpha$-particles, 290~MeV/n Be, B and C ions, and 200, 400 and 600~keV Kr ions in fine grain nuclear emulsion as a function of LET (Fig.~4 in Ref.~\cite{emulsion_NATSUME}).
		The electronic stopping power at incident energy were used for Kr ions (green triangle) \cite{emulsion_NATSUME}.
		Circles for Kr are replotted at averaged $LET_{\rm el}$ (blue circle; present work, see the text).
		Open and closed symbols show the difference in developer.
	}
	\label{fig_5}
\end{figure}
\par
The mean hit density, $(n_{\rm F}-1)/R$, where $n_{\rm F}$ is the number of the filaments, for $\alpha$-particles, 290 MeV/n relativistic Be, B and C ions, and low velocity Kr ions in fine grain nuclear emulsion reported are shown as a function of $LET$ in Fig.~\ref{fig_5} (Fig.~4 in Ref.~\cite{emulsion_NATSUME}).
The grain size is 40~nm.
Figure~\ref{fig_5} demonstrates an importance of $LET_{\rm el}$.
The electronic stopping power \cite{SRIM2010} at incident energy were used for Kr ions (closed triangle) \cite{emulsion_NATSUME}.
Points for $\alpha$-particles and relativistic ions are on a straight line (broken line) in a log-log plot.
Whereas, the data for Kr ions are almost constant and do not stay on the line.
The electronic stopping power for 200, 400 and 600 keV Kr ions changes about a factor $\sim$2.5; however, the hit density is within the experimental error.
The points for Kr ions were replotted also at $\langle LET_{\rm el}\rangle$ in Fig.~\ref{fig_5}.
The points came near to the straight line, when the difference in developer is taken into account (solid line).
$\langle LET_{\rm el}\rangle$ for those Kr ions differ less than 25\%.
The stopping power describes how the incident ion losses its energy and does not take secondary effects into account.
LET, on the other hand, describes the energy deposited on the target material.
The energy deposition due to secondary particles are also considered.

\subsection{Track structure}
\begin{figure}[t]
	\includegraphics[width=0.9\linewidth]{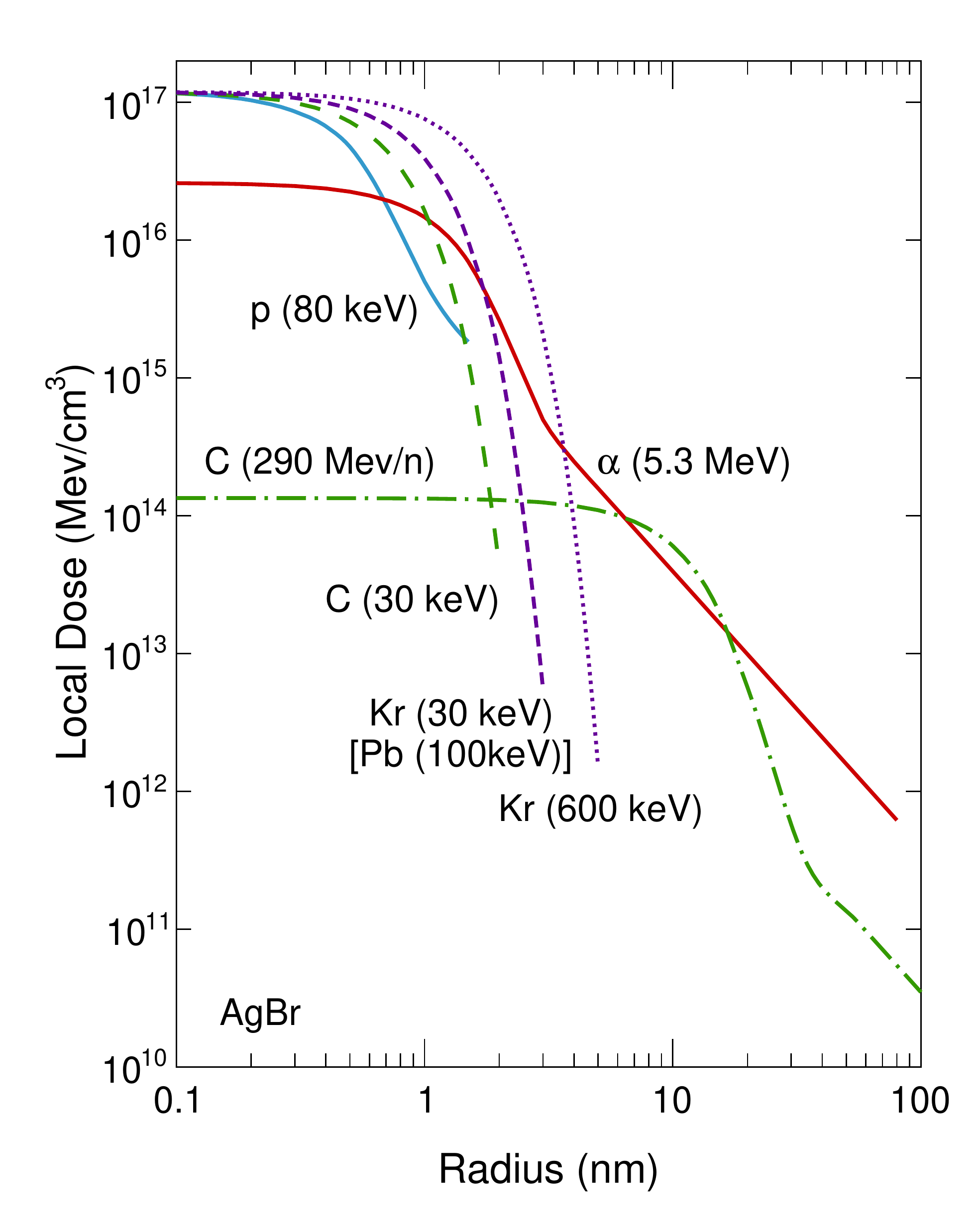}
	\caption{
		Initial radial distribution of dose in AgBr crystal due to various ions showing the core and penumbra of heavy-ion track structure.
		A part of penumbra is shown for 290~MeV/n C ions.
		The radial distributions for 290~MeV/n Be and B ions are the same as that for C ions when difference in LET is taken into account, therefore, Be and B ions are not shown.
		Curves for slow C and Kr ions have only undifferentiated core.
		The initial radial distribution for 100~keV Pb is practically the same as 30~keV Kr ions. 
	}
	\label{fig_6}
\end{figure}

For the directional detection of WIMPs, the information on LET dependence of hit density is not sufficient.
The initial radial distributions of local dose for various particles in AgBr crystal are estimated and compared in Fig.~\ref{fig_6} for further studies.
The averaged value was taken for LET as in Fig.~\ref{fig_5} and the initial energy was taken for $r_{\rm c}$ and $r_{\rm p}$, following custom.
Values of $r_{\rm c}$ and $r_{\rm p}$ were calculated for AgBr crystal.
The range of $\delta$-rays for fast ions is larger than the grain size.
However, the range in AgBr, not in emulsion, were taken to obtain $r_{\rm p}$.
This is because the $r_{\rm c}/r_{\rm p}$ ratio determines the core/penumbra ratio and consequently the core density.
The core radius is smaller than the grain size.
The core density is much important than that in the penumbra in understanding the response of emulsion for WIMP searches.
The values for $r_{\rm p}$ was obtained by dividing the range (in g/cm$^2$) for the maximum energy $E_{\delta}^{\rm max}$ of $\delta$-rays in emulsion \cite{WNB1963} by the density for AgBr for $E_{\delta}^{\rm max} > 10$ keV.
At a low energy, extrapolated ranges given by `best-fit' expressions, Eq.~(\ref{eq_Rex}), by Iskef et al. \cite{Iskef1983} were taken.
\par
The dose of penumbra decreases as $r^2$ at a large $r$; however, it should be noted that this is the averaged value.
The penumbra consists of $\delta$-rays, therefore, the local LET should be regarded as that of $\delta$-rays.
The dose in the core for relativistic particles is more than two orders of magnitude lower than that for $\alpha$-particles, however, low LET core is not homogeneous, spar and blob formation has to be considered \cite{Mozumder1999}.
\par
The track of slow recoil ions consists only of an undifferentiated core.
The core radius depends only on $LET_{\rm el}$ for slow recoil ions when core expansion is taken.
Most of energy is deposited within the radius of the grain size as can be seen in Fig.~\ref{fig_6}, therefore, is likely to stay within the grain if it is recoiled in the grain.
\par
$\alpha$-particles can be a good measure to simulating the radiation effects for recoil ions for many WIMP detectors since the track core structure and density are alike to each other \cite{DM_search4,Hitachi2005}.
The energy of Ag and Br recoil ions, which the directional searches aim at, are generally higher than those for non-directional searches.
$LET_{\rm el}$ for recoil ions are considerably larger than those for $\alpha$-particles as shown in Fig.~\ref{fig_5}.
In addition, the core density calculated for recoil ions are much higher than that for $\alpha$-particles in AgBr crystal as shown in Fig.~\ref{fig_6}.
Having the band structure, $E_1$ ($E_{\rm g}$) and $W$ for AgBr crystal are much smaller than those for atoms and molecules without the band structure.
Those makes relatively longer $r_{\rm c}$ and less dense core for $\alpha$-particles
The core radius for recoil ions and $\alpha$-particles are similar in magnitude; however, the local dose at the center of the core for recoil ions are about 5 times that for $\alpha$-particles.
The effects of the difference in LET and local dose are to be investigated.

\subsection{Background}
The very heavy recoil ions in $\alpha$-decay may leave WIMP-like tracks as mentioned in section \ref{sec_very_heavy_recoil_ions_in_alpha_decay}.
The values of $q_{\rm nc}$ estimated for 100 -- 180~keV Pb ions in AgBr crystal are shown in Fig.~\ref{fig_2}.
Value of $\langle LET_{\rm el}\rangle$ calculated for recoil ions and Pb ions are quite close to each other as shown in Table~\ref{tab:table1}.
In fact, the $\langle LET_{\rm el}\rangle$ values for 100~keV and 170~keV Pb ions are 360 and 540~keV/$\mu$m, respectively, and those for 30~keV and 100~keV Kr ions are 340 and 550~keV/$\mu$m, respectively.
The initial radial distribution for 30~keV Kr ions and 100~keV Pb ions are practically the same in Fig.~\ref{fig_6}.
Usually, one does not see the heavy recoil ions and $\alpha$-particles as separate particles, since they are produced at the same time.
The Pb track is associated with the much larger $\alpha$-particle track.
However, in some cases, Pb recoiled at the boundary of AgBr crystal and gelatin such that an $\alpha$-particle goes into gelatin or escapes from the emulsion, and the Pb recoil fully goes into the crystal.
Then, the recoil Pb ion can produce a WIMP-like signal. 
\par
Search for dark matter requires large exposure, i.e., mass $\times$ time, therefore to obtain a good signal to noise ratio, it is better to reject causes of noises as much as possible by adjusting sensitivity of emulsion.
Major contribution to the background in underground laboratories are $\gamma$-rays and neutrons.
Since $LET$ for the electron is much smaller than $LET_{\rm el}$ for recoil ions, $\gamma$-rays may be disregarded by setting sensitivity of emulsion and/or by cryogenic crystal effect\cite{Kimura2017}.
Neutrons can produce WIMP-like signals as the neutron scattering is used to produce recoil ions to mock dark matter signal in detector media \cite{emulsion_NAKA}.
These signals are to be distinguished using daily modulation by directionality measurements.
The directional detector usually sits on an equatorial mount.
The position and direction will give some ways to reject the very heavy recoil ions in $\alpha$-decay as well as $\gamma$-rays and neutrons.
In addition to these, a special caution should be payed to the knock-on protons produced by fast neutron \cite{Park2013} recoiling the hydrogen in the gelatin of the emulsion, as used in film badges, produce latent image in AgBr crystal.
The result for 5.3~MeV $\alpha$-particles shown in Fig.~\ref{fig_6} can also be interpreted as $\sim$1.3~MeV protons except $LET$ value is different.
The difference in $LET$ can be taken care of simply, since Bethe formula, for stopping power for fast ions, scales as $Z_1^2$, a quarter $LET$ for $\alpha$-particles will give the distribution for $\sim$1.3~MeV protons.
Fast protons may be disregarded from its range.
However, for directional capability, at least 2 -- 3 grains are necessary, this makes 100 -- 200~nm.
With usual optical reading system, submicron track length will be necessary.
Protons of energy less than $\sim$50~keV becomes difficult to distinguish from WIMP signal by means of the range alone.
The energy of the Bragg peak for protons in Ag Br crystal is at about 80~keV.
The range for 80~keV proton is 0.74~$\mu$m.
The radius of the core $r_{\rm c}$ given by the Bohr-criterion becomes smaller than $a$ at about 70~keV for protons, then $r_{\rm c}$ = $a$ = 0.288~nm is regarded as the minimum core radius.
The result for 80~keV protons shows dose distribution of almost minimum radius.
The maximum local dose calculated for 80~keV protons using $r_{\rm c}$ = 0.3~nm exceeds $D_{\rm max}$, therefore, the core expansion, $r_{\rm ex}$ = 0.46~nm, was taken.
Initial radial distribution of dose in AgBr crystal due to 80~keV protons is shown in Fig.~\ref{fig_6}.
The local dose for penumbra due to $\delta$-rays ($r > r_{\rm ex}$) was given by Eq.~(\ref{eq_Dk2}) with $r_{\rm c}$ = 0.3~nm and $r_{\rm p}$ = 1.5~nm.
The local dose for penumbra may not play an important role for low energy protons.
It is practically the same if the track consists only of an undifferentiated core of $r_{\rm ex}$ = 0.54~nm.
An undifferentiated core of $r_{\rm ex}$ = 0.48~nm was taken for 25~keV protons in Table \ref{tab:table1}.

\subsection{General remarks}
The errors in $q_{\rm nc}$ may be 5 -- 15\% in the Lindhard model as discussed in Ref.~\cite{Hitachi2005}.
We took $k = 0.15$ instead of $k = 0.158$ for Kr ions in Kr.
This simplification may underestimate $\eta$ value about 4\%.
The errors in the independent element approximation for C in AgBr may be 10 -- 20\%.
Overall uncertainties in the present calculation can be considerably larger.
However, the latent image production mechanism is quite complicated and quantitative prediction is very difficult because blackness depends on many factors.
The relative values in $q_{\rm nc}$, $LET_{\rm el}$ and the track parameters will do and errors in the relative values are much smaller. 
\par
It is important to know if the latent image formation is determined by LET (or energy per crystal) or local dose (local deposited-energy density).
$\langle LET\rangle$ for protons are about 100~keV/$\mu$m and are much smaller than $\langle LET_{\rm el}\rangle$ for Kr recoil ions and 1/3 that for C recoil ions in a submicron range.
If LET is the main factor, it may not be difficult to disregard protons from Kr recoil ions.
It may be harder for C recoils, however, it may still be possible to reject protons.
However, the maximum local dose for protons are the same as those for C recoil ions and Kr recoil ions.
It may be naive to assume that the grain becomes developable when the local dose shown in Fig.~\ref{fig_6} exceeds a particular threshold value and the dose above this threshold will contribute the blackness of the track and determine the sensitivity.
The track core for heavy ions is very thin therefore it can rapidly diffuse out for the radial direction.
The reaction kinetics may have to be considered \cite{Hitachi1992}.
\par
The nuclear stopping process recoils Ag and Br atoms in the crystal; the replacement of atom may cause the distortion of crystal.
The effect was not treated here and have to be considered.
The energy $E_{\rm th}$ spent as heat may increase the local temperature of the crystal and can influence the latent image formation.
The energy $\nu$ goes to thermal energy.
The thermal energy produces phonon and can be used \cite{Kimura2017}.
The nuclear LET ($LET_{\rm nc}$), the energy given to nuclear motion in the stopping process per unit path length, is expressed as
\begin{equation}\label{eq_LETnc}
\begin{split}
LET_{\rm nc}&=-\frac{{\rm d}E_\nu}{{\rm d}x}=-\frac{{\rm d}E_\nu}{{\rm d}E} \frac{{\rm d}E}{{\rm d}x}=\frac{{\rm d}E_\nu}{{\rm d}E} S_{\rm T} \\ &\approx\frac{\Delta E_\nu}{\Delta E} S_{\rm T},
\end{split}
\end{equation}
In addition, some parts of $\eta$ also contributes to thermal energy.
A part of $\eta$ is used as light emission \cite{Moser1971} and another is used as chemical reactions; the rest is spent as heat in emulsion.
Slow ions suffer significant deviation from the initial trajectory due to scattering.
The track detours and has branches.
These effects also have to be considered.
\par
The present model is simple and assumptions are clear.
One can refine the calculation or extent the model when needed.
The track structure obtained here do not immediately predict the latent image in emulsion because of the complex nature of response of emulsion to ionizing radiation.
The present results can be used for adjusting sensitivity, grain size, finding optimum developing conditions, etc.

\section{Summary}
The electronic energy deposition due to slow C and Kr ions in nuclear emulsion was estimated for directional detection of dark matter (WIMPs).
The concept of electronic LET has been introduced and its importance were shown to explain the mean-hit density for slow Kr ions, $\alpha$-particles and relativistic ions.
The so-called core and penumbra of heavy-ion track structure is considered and modified for various ions.
The initial radial distributions of electronic dose for various ions are presented and compared for further studies.
The tracks due to very heavy recoil ions, 100 -- 180~keV Pb ions, produced in $\alpha$-decay are also estimated.
The track for protons was also studied to evaluate influence of neutrons which is one of main background.
It is demonstrated that some backgrounds can be difficult to distinguish with WIMP signals by difference in LET or in the track structure, in such cases directional detection becomes important.

\begin{acknowledgements}
We would like to thank Prof. T. Tani for valuable discussion and Dr. T. Naka for providing information on nuclear emulsion.
The work described herein was supported in part by the Office of Basic Energy Science of the Department of Energy. 
This is document number NDRL 5242 from the Notre Dame Radiation Laboratory.
\end{acknowledgements}

\end{document}